# Analytical solution of Stokes flow inside an evaporating sessile drop: Spherical and cylindrical cap shapes

Hassan Masoud and James D. Felske

*Department of Mechanical and Aerospace Engineering, State University of New York at Buffalo, Buffalo, New York 14260, USA*

Exact analytical solutions are derived for the Stokes flows within evaporating sessile drops of spherical and cylindrical cap shapes. The results are valid for arbitrary contact angle. Solutions are obtained for arbitrary evaporative flux distributions along the free surface as long as the flux is bounded at the contact line. The field equations, $E^4\psi = 0$ and $\nabla^4\psi = 0$, are solved for the spherical and cylindrical cap cases, respectively. Specific results and computations are presented for evaporation corresponding to uniform flux and to purely diffusive gas phase transport into an infinite ambient. Wetting and non-wetting contact angles are considered with the flow patterns in each case being illustrated. For the spherical cap with evaporation controlled by vapor phase diffusion, when the contact angle lies in the range $0 \leq \theta_c < \pi/2$, the mass flux of vapor becomes singular at the contact line. This condition required modification when solving for the liquid phase transport. Droplets in all of the above categories are considered for the following two cases: the contact lines are either pinned or free to move during evaporation. The present viscous flow behavior is compared to the inviscid flow behavior previously reported. It is seen that the streamlines for viscous flow lie farther from the substrate than the corresponding inviscid ones.

# I. INTRODUCTION

When a sessile droplet containing solid constituents dries on a substrate, the solute particles deposit on the surface in various patterns. For example, a particular deposition pattern - the ring - has important implications for: DNA mapping,[1, 2] ink-jet printing,[3-5] production of crystals,[6-8] and coatings.

The deposition pattern produced depends upon the flow within the drop. As shown previously,[9] the flow pattern strongly depends on the combination of evaporative flux, shape of the free surface, and behavior of the contact line. For instance, flow can be toward, away or both toward and away from the contact line under different combinations of these factors.

Flow inside an evaporating sessile droplet has been studied analytically, semi-analytically and numerically. Deegan[10-12] and Popov[13] utilized the vertically averaged velocity in considering the behavior of velocities at small contact angles. Solutions in the limit of lubrication theory were obtained numerically by Fischer[14] and semi-analytically by Hu and Larson.[15] Stokes flow for contact angles ranging from zero to ninety degrees was determined numerically by Hu and Larson[15] and Widjaja et al.[16] The exact analytical solution for irrotational flow was derived for hemispheres by Tarasevich,[17] for spherical caps by Masoud and Felske,[9] and for hemicylinders and cylindrical caps by Petsi and Burganos.[18, 19]

The present study derives the exact analytical solutions for the Stokes flows within evaporating, sessile, spherical and cylindrical cap drops. Arbitrary contact angles ($0 \leq \theta_c \leq \pi$) are considered as well as arbitrary distributions of evaporative flux along the free surface. The contact line of the droplet is allowed to be either pinned or free to move during evaporation. Analytical expressions for the expansion coefficients are given for the case of uniform flux. To the best of our knowledge, the Stokes flow within an evaporating, sessile, non-wetting drop has not been reported by any approach, analytical or numerical.

## II.  MODEL DEVELOPMENT AND SOLUTIONS

The characteristic length of droplets under consideration, the distance from the axis of symmetry to the contact line ($R$), are $\sim 1\,mm$. For evaporation under room conditions, the characteristic velocity in a water drop of this size is $\sim 1\,\mu m/s$, and the Reynolds number is $\sim 10^{-3}$.[15] Consequently, for this type of flow, the inertial forces are negligible (Stokes flow) and the continuity and momentum equations become, respectively:

$$\nabla.\mathbf{V} = 0 , \tag{1}$$

and

$$\mu \nabla^2 \mathbf{V} = \nabla p . \tag{2}$$

Taking the curl of Eq. (2),

$$\nabla \times (\nabla^2 \mathbf{V}) = 0 , \tag{3}$$

and introducing the vorticity, $\boldsymbol{\omega} = \nabla \times \mathbf{V}$, Eq. (3) may be written as,

$$\nabla^2 \boldsymbol{\omega} = 0 . \tag{4}$$

The application of the above to the two geometries of interest will now be considered.

## A.  Spherical cap shape

### 1.  Geometry

The shape acquired by a drop on a substrate is determined by surface tension forces on the free surface and pressure forces within the drop. The pressure within the drop varies from its value at the interface as a result of flow within the drop and the force of gravity. The pressure in the liquid at the interface equals the pressure jump across the curved interface. The magnitude of this jump varies with the local curvature. The Bond number is the ratio of gravitational to surface tension forces. The capillary number expresses the ratio of viscous to surface tension forces. For millimeter size drops the capillary number is small[15] and the bond number is considerably less than one.[17] Since both the capillary and Bond numbers are very small, surface tension dominates and, hence, the droplet takes the shape of a spherical cap.

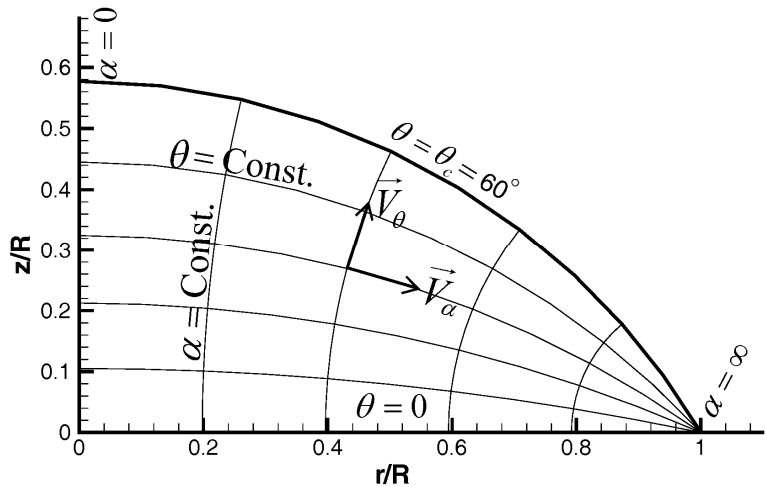

FIG. 1. Toroidal coordinates: lines of constant $\alpha$ and $\theta$; positive velocity components.

The drop geometry is that of a sphere cut by a plane. The natural coordinates are therefore toroidal – see Fig. 1. Since the shape and the boundary conditions are azimuthally independent, so is the flow within the drop. Figure 1 shows a cross section of the toroidal geometry at a given azimuthal angle, $\varphi$. The metric coefficients for the this geometry are:

$$h_\alpha = h_\theta = h_\varphi / \sinh\alpha = R(\cosh\alpha + \cos\theta)^{-1}, \qquad (5)$$

where $0 \le \alpha \le \infty$, $-\pi \le \theta \le \pi$ and $0 \le \varphi \le 2\pi$. It is noted here that for the spherical cap, the required cylindrical coordinates are related to the toroidal coordinates by:

$$r / \sinh\alpha = z / \sin\theta = R(\cosh\alpha + \cos\theta)^{-1}. \qquad (6)$$

## 2. Field equation

For 2D and 3D axi-symmetric flows, velocity components may be expressed in terms of a stream function. The stream function, $\psi$, is defined to satisfiy the continuity equation:

$$V_\alpha = \frac{1}{h_\theta h_\varphi} \frac{\partial\psi}{\partial\theta} = \frac{(\cosh\alpha + \cos\theta)^2}{R^2 \sinh\alpha} \frac{\partial\psi}{\partial\theta}, \qquad (7)$$

$$V_\theta = -\frac{1}{h_\varphi h_\alpha} \frac{\partial\psi}{\partial\alpha} = -\frac{(\cosh\alpha + \cos\theta)^2}{R^2 \sinh\alpha} \frac{\partial\psi}{\partial\alpha}. \qquad (8)$$

The field equation for $\psi$ follows from writing Eq. (4) for axially-symmetric flow. Substituting from above for $V_\alpha$ and $V_\theta$ in terms of $\psi$, the vorticity is generally related to the stream function by:

$$\boldsymbol{\omega} = (\hat{\mathbf{e}}_\varphi / h_\varphi) E^2 \psi, \qquad (9)$$

where,

$$E^2\psi = \frac{\sinh\alpha\,(\cosh\alpha + \cos\theta)}{R^2} \left[ \frac{\partial}{\partial\alpha}(\frac{\cosh\alpha + \cos\theta}{\sinh\alpha} \frac{\partial\psi}{\partial\alpha}) + \frac{\partial}{\partial\theta}(\frac{\cosh\alpha + \cos\theta}{\sinh\alpha} \frac{\partial\psi}{\partial\theta}) \right].$$
$$(10)$$

Using the vector identities $\nabla^2\boldsymbol{\omega} = \nabla(\nabla\cdot\boldsymbol{\omega}) - \nabla\times(\nabla\times\boldsymbol{\omega})$ and $\nabla\cdot(\nabla\times\mathbf{V}) = 0$, Eq. (4) may be written as

$$\nabla \times (\nabla \times \boldsymbol{\omega}) = 0. \tag{11}$$

Repeated application of the curl ($\nabla \times$) operator to Eq. (9) results in,

$$\nabla \times (\nabla \times \boldsymbol{\omega}) = -(\hat{\mathbf{e}}_\varphi / h_\varphi) E^2 (E^2 \psi). \tag{12}$$

Consequently, the field equation for $\psi$ which corresponds to $\nabla^2 \boldsymbol{\omega} = 0$, is

$$E^4 \psi = 0. \tag{13}$$

### 3. *Integration of* $E^4 \psi = 0$

An analytical solution to $E^4 \psi = 0$ subject to the appropriate boundary conditions is obtainable by the method of separation of variables. The solution is discussed below.

### (a) *Separation of variables.*

Following Khuri & Waswaz,[20] a separable solution to $E^4 \psi = 0$ in toroidal coordinates begins from assuming the solution to be of the form

$$\psi(\alpha, \theta) = (\cosh \alpha + \cos \theta)^{-3/2} \phi(\alpha, \theta). \tag{14}$$

Substituting the above into Eq. (10) and then repeating the $E^2$ operation leads to

$$\frac{\partial^4 \phi}{\partial \alpha^4} + 2 \frac{\partial^4 \phi}{\partial \alpha^2 \partial \theta^2} + \frac{\partial^4 \phi}{\partial \theta^4} - 2 \coth \alpha \frac{\partial^3 \phi}{\partial \alpha^3} - 2 \coth \alpha \frac{\partial^3 \phi}{\partial \alpha \partial \theta^2} + (3 \coth^2 \alpha - 3.5) \frac{\partial^2 \phi}{\partial \alpha^2}$$

$$+ (5/4) \frac{\partial^2 \phi}{\partial \theta^2} + \coth \alpha (4.5 - 3 \coth^2 \alpha) \frac{\partial \phi}{\partial \alpha} + (9/16) \phi = 0. \tag{15}$$

Assuming a product form for separating variables in Eq. (15),

$$\phi(\alpha, \theta) = f(\alpha) g(\theta), \tag{16}$$

leads to

$$g^{(4)} + 2\left[\frac{f^{(2)}}{f} - \coth\alpha\frac{f^{(1)}}{f} + 5/4\right]g^{(2)} + \left[\frac{f^{(4)}}{f} - 2\coth\alpha\frac{f^{(3)}}{f}\right.$$

$$\left. + (3\coth^2\alpha - 3.5)\frac{f^{(2)}}{f} + \coth\alpha(4.5 - 3\coth^2\alpha)\frac{f^{(1)}}{f} + 9/16\right]g = 0. \tag{17}$$

To be consistent with the assumed separable form, Eq. (16), the two bracketed coefficients in Eq. (17) must be constants. Denoting the constant for the first bracket as $(1-\tau^2)$ (in order to obtain an established differential equation for the eigenfunctions in $\alpha$), the Gegenbauer equation is obtained:

$$f^{(2)} - \coth\alpha\, f^{(1)} + (\tau^2 + 1/4)\, f = 0. \tag{18}$$

The solution to Eq. (18) is

$$f(\alpha) = A(\tau)C_{1/2+i\tau}^{-1/2}(\cosh\alpha) + B(\tau)C_{1/2+i\tau}^{*-1/2}(\cosh\alpha), \tag{19}$$

where, $C_{1/2+i\tau}^{-1/2}$ and $C_{1/2+i\tau}^{*-1/2}$ are Gegenbauer functions of the first and second kind, of order -1/2. Using Eq. (18) and its first and second derivatives, the second bracketed coefficient in Eq. (17) is readily shown to be:

$$\frac{f^{(4)}}{f} - 2\coth\alpha\frac{f^{(3)}}{f} + (3\coth^2\alpha - 3.5)\frac{f^{(2)}}{f} + \coth\alpha(4.5 - 3\coth^2\alpha)\frac{f^{(1)}}{f} + 9/16$$

$$= (\tau^2 + 1)^2. \tag{20}$$

Inserting the above constant values for the first and second brackets, Eq. (17) becomes:

$$\frac{d^4g}{d\theta^4} + 2(1-\tau^2)\frac{d^2g}{d\theta^2} + (\tau^2 + 1)^2 g = 0. \tag{21}$$

The solution to this constant coefficient equation follows directly and is given by

$$g(\theta) = \sinh(\tau\theta)\big[C(\tau)\sin\theta + D(\tau)\cos\theta\big] + \cosh(\tau\theta)\big[E(\tau)\sin\theta + F(\tau)\cos\theta\big], \tag{22}$$

where, $C(\tau)$, $D(\tau)$, $E(\tau)$, and $F(\tau)$ are real functions. Since $0 \le \alpha < \infty$, the associated eigenvalues are continuously distributed: $0 \le \tau < \infty$. The general solution is then:

$$\psi(\alpha,\theta) = (\cosh\alpha + \cos\theta)^{-3/2} \int_0^\infty [A(\tau)C_{1/2+i\tau}^{-1/2}(\cosh\alpha) + B(\tau)C_{1/2+i\tau}^{*-1/2}(\cosh\alpha)]$$

$$\times \left\{ \sinh(\tau\theta)\big[C(\tau)\sin\theta + D(\tau)\cos\theta\big] \right.$$

$$\left. + \cosh(\tau\theta)\big[E(\tau)\sin\theta + F(\tau)\cos\theta\big] \right\} d\tau . \qquad (23)$$

*(b) Boundary conditions.*

To establish boundary conditions on the stream function, $\psi$, it is first noted that the velocity components normal to the axis of symmetry and normal to the solid surface vanish: $V_\alpha(0,\theta) = 0$ and $V_\theta(\alpha,0) = 0$. Hence, the stream function is constant along the axis of symmetry and along the solid surface. Since these surfaces intersect, the constant is the same for both surfaces. Because its value does not affect the predicted velocities, it is arbitrarily set to zero. The no slip condition requires that the velocity component tangent to the solid surface vanish at the surface, $V_\alpha(\alpha,0) = 0$. Along the axis of symmetry, derivatives normal to the axis are also zero; in particular $\partial V_\theta / \partial\alpha(0,\theta) = 0$. The corresponding boundary conditions are:

$$\psi(0,\theta) = 0, \qquad (24)$$

$$\psi(\alpha,0) = 0, \qquad (25)$$

$$\left. \frac{\partial\psi}{\partial\theta} \right|_{\theta=0} = 0, \qquad (26)$$

and

$$\frac{\partial}{\partial \alpha}\left[\frac{(\cosh\alpha + \cos\theta)^2}{\sinh\alpha}\frac{\partial\psi}{\partial\alpha}\right]_{\alpha = 0} = 0. \tag{27}$$

Applying these conditions to Eq. (23) the following are obtained: from Eq. (24), $B(\tau) = 0$;[9] from Eq. (25)

$$F(\tau) = 0, \tag{28}$$

and, from Eq. (26),

$$E(\tau) + \tau D(\tau) = 0. \tag{29}$$

Incorporating these into Eq. (23), the stream function becomes:

$$\psi(\alpha,\theta) = (\cosh\alpha + \cos\theta)^{-3/2}\int_0^\infty K(\theta,\tau)C_{1/2+i\tau}^{-1/2}(\cosh\alpha)\,d\tau, \tag{30}$$

where,

$$K(\theta,\tau) = k_1(\tau)\sin\theta\sinh(\tau\theta) + k_2(\tau)\big[\cos\theta\sinh(\tau\theta) - \tau\sin\theta\cosh(\tau\theta)\big]. \tag{31}$$

The third and fourth boundary conditions in the $\alpha$-direction require the velocity components to be non-singular at the contact line:

$$\frac{(\cosh\alpha + \cos\theta)^2}{\sinh\alpha}\frac{\partial\psi}{\partial\theta}(\infty,\theta) = \text{finite} \tag{32}$$

and

$$\frac{(\cosh\alpha + \cos\theta)^2}{\sinh\alpha}\frac{\partial\psi}{\partial\alpha}(\infty,\theta) = \text{finite}. \tag{33}$$

These two conditions and Eq. (27) are satisfied for physically meaningful distributions of evaporative flux at the free surface.

The unknown coefficients, $k_1(\tau)$ and, $k_2(\tau)$ can be obtained from the distribution of the stream function at the free surface in conjunction with the zero shear stress boundary condition, $\tau_{\alpha\theta}(\alpha, \theta_c) = 0$. $K(\theta_c, \tau)$ can be written in terms of the stream function at the free surface using the integral transform presented in Ref. 9,

$$K(\theta_c, \tau) = \tau(\tau^2 + 1/4)\tanh(\pi\tau)\int_0^\infty \frac{\psi(\alpha, \theta_c)(\cosh\alpha + \cos\theta_c)^{3/2}}{\sinh\alpha} C_{1/2+i\tau}^{-1/2}(\cosh\alpha)\, d\alpha \,. \quad (34)$$

Using Eqs. (4-14.1), (4-14.2) and (A-7.7) of Ref. 21, vanishing of the shear stress at the free surface gives

$$\tau_{\alpha\theta}\big|_{\theta=\theta_c} = -\mu\left\{ \frac{\partial}{\partial\alpha}\big[(\cosh\alpha + \cos\theta)(V_\theta/R)\big] + \frac{\partial}{\partial\theta}\big[(\cosh\alpha + \cos\theta)(V_\alpha/R)\big] \right\}_{\theta=\theta_c}$$

$$= 0\,. \quad (35)$$

Therefore,

$$\widetilde{K}(\theta_c, \tau) = \tau(\tau^2 + 1/4)\tanh(\pi\tau)\int_0^\infty \frac{\widetilde{\psi}(\alpha, \theta_c)}{\sinh\alpha} C_{1/2+i\tau}^{-1/2}(\cosh\alpha)\, d\alpha \,, \quad (36)$$

in which,

$$\widetilde{\psi}(\alpha, \theta_c) = -(\cosh\alpha + \cos\theta_c)^{-1/2}\left\{ \frac{R^2\sinh\alpha}{\cosh\alpha + \cos\theta_c} \frac{\partial}{\partial\alpha}\big[(\cosh\alpha + \cos\theta_c)V_\theta(\alpha, \theta_c)\big] \right.$$

$$\left. + \frac{3\psi(\alpha, \theta_c)}{2}\big[\cos\theta_c(\cosh\alpha + \cos\theta_c) - \sin^2\theta_c/2\big] \right\}$$

$$\quad (37)$$

and

$$\widetilde{K}(\theta_c, \tau) = \frac{\partial^2}{\partial^2\theta} K(\theta_c, \tau)$$

$$= k_1(\tau)\big[(\tau^2 - 1)\sin\theta_c\sinh(\tau\theta_c) + 2\tau\cos\theta_c\cosh(\tau\theta_c)\big]$$

$$- k_2(\tau)(\tau^2 + 1)\big[\cos\theta_c\sinh(\tau\theta_c) + \tau\sin\theta_c\cosh(\tau\theta_c)\big]\,. \quad (38)$$

Using Eqs. (34) and (36), the coefficients, $k_1(\tau)$ and $k_1(\tau)$, are found to be

$$k_1(\tau) = \frac{N_2(\tau,\theta_c)K(\tau,\theta_c) + N_1(\tau;\theta_c)\widetilde{K}(\tau,\theta_c)}{N_2(\tau,\theta_c)M_1(\tau,\theta_c) + N_1(\tau,\theta_c)M_2(\tau,\theta_c)}, \tag{39}$$

and

$$k_2(\tau) = \frac{M_2(\tau,\theta_c)K(\tau,\theta_c) - M_1(\tau,\theta_c)\widetilde{K}(\tau,\theta_c)}{N_2(\tau,\theta_c)M_1(\tau,\theta_c) + N_1(\tau,\theta_c)M_2(\tau,\theta_c)}, \tag{40}$$

where,

$$M_1(\tau,\theta_c) = \sin\theta_c\sinh(\tau\theta_c), \tag{41}$$

$$M_2(\tau,\theta_c) = (\tau^2-1)\sin\theta_c\sinh(\tau\theta_c) + 2\tau\cos\theta_c\cosh(\tau\theta_c), \tag{42}$$

$$N_1(\tau,\theta_c) = \cos\theta\sinh(\tau\theta) - \tau\sin\theta\cosh(\tau\theta), \tag{43}$$

and

$$N_2(\tau,\theta_c) = (\tau^2+1)\left[\cos\theta_c\sinh(\tau\theta_c) + \tau\sin\theta_c\cosh(\tau\theta_c)\right]. \tag{44}$$

The final boundary condition is the known distribution of evaporative flux along the free surface. This distribution is determined by the gas phase mass transport and is independent of the flow within the drop. The velocity and stream function at the free surface follow directly from this flux. The velocity is obtained from mass conservation whereby the mass flux in the liquid equals the mass flux in the gas. In terms of the liquid, the evaporation rate is:

$$J(\alpha) = \rho\left[V_\theta(\alpha,\theta_c) - V_{\theta,B}(\alpha)\right], \tag{45}$$

where, $J(\alpha)$ is the evaporative flux at the free surface, $\rho$ is the liquid density and $V_{\theta,B}$ is the speed at which the boundary is moving in the direction normal to itself. From Eq. (8), the boundary condition on $\psi$ may then be written in terms of $V_\theta(\alpha,\theta)\big|_{\theta=\theta_c}$:

$$\psi(\alpha,\theta_c) = -\int_0^\alpha \frac{R^2\sinh\alpha'}{(\cosh\alpha' + \cos\theta_c)^2}\,V_\theta(\alpha',\theta_c)\,d\alpha'$$

$$= -\int_0^\alpha \frac{R^2 \sinh \alpha'}{(\cosh \alpha' + \cos \theta_c)^2} \left[ J(\alpha') / \rho + V_{\theta,B}(\alpha') \right] d\alpha'. \tag{46}$$

Finally, from $\psi(\alpha, \theta)$, the velocity distribution may be calculated. The velocity components in toroidal coordinates follow from Eqs. (7) and (8):

$$V_\alpha(\alpha, \theta) = \frac{\sqrt{\cosh \alpha + \cos \theta}}{R^2 \sinh \alpha} \Bigg[ (3 \sin \theta / 2) \sqrt{\cosh \alpha + \cos \theta} \, \psi(\alpha, \theta)$$

$$+ \int_0^\infty \frac{\partial K(\theta, \tau)}{\partial \theta} C_{1/2+i\tau}^{-1/2} (\cosh \alpha) \, d\tau \Bigg], \tag{47}$$

$$V_\theta(\alpha, \theta) = \frac{\sqrt{\cosh \alpha + \cos \theta}}{R^2} \Bigg\{ \frac{3 \sqrt{\cosh \alpha + \cos \theta} \, \psi(\alpha, \theta)}{2}$$

$$+ \int_0^\infty K(\theta, \tau) P_{-1/2+i\tau} (\cosh \alpha) \, d\tau \Bigg\}. \tag{48}$$

where, $P_{-1/2+i\tau}(x)$ is the conical function of the first kind. It is noted that the velocity components in cylindrical coordinates are useful for visualizing and physically interpreting the flow field. The radial and axial components of the velocity in these coordinates may be calculated from Eqs. (50) and (51) of Ref. 9.

## B. Cylindrical cap shape

### 1. Geometry

Similar to spherical drops, the sizes of 2D drops (liquid lines) being considered are small enough that surface tension is the dominant force defining their cross-sectional shape. Consequently, they have the cylindrical cap geometry whose free surface is exactly mapped in bipolar coordinates. Therefore, analysis proceeds in this coordinate system. Bipolar coordinates $(\alpha, \theta)$ are shown in Fig. 2 along with Cartesian coordinates $(x, y)$. (Note that the cross-section of a cylindrical cap in bipolar coordinates is identical to the cross-section of a spherical cap in toroidal coordinates – Fig. 1.) The metric coefficients for bipolar geometry are:

$$h_\alpha = h_\theta = R(\cosh \alpha + \cos \theta)^{-1},\tag{49}$$

where, $-\infty \le \alpha \le \infty$ and $-\pi \le \theta \le \pi$. Bipolar and Cartesian coordinates are related by:

$$x / \sinh \alpha = y / \sin \theta = R(\cosh \alpha + \cos \theta)^{-1}.\tag{50}$$

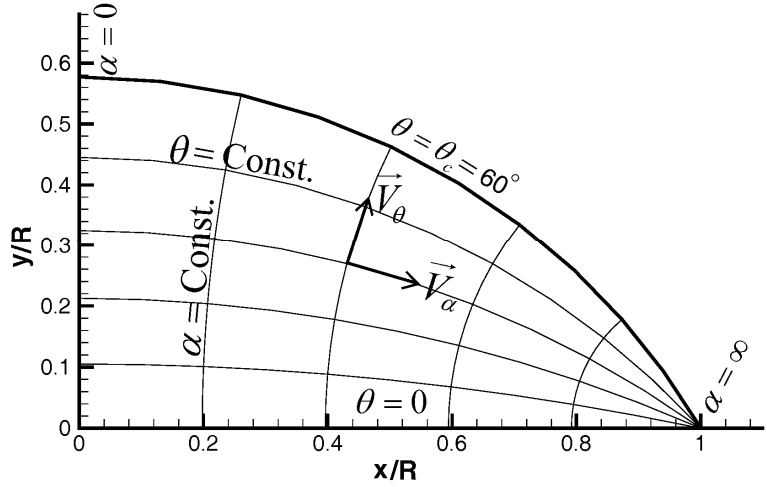

FIG. 2. Bipolar coordinates: lines of constant $\alpha$ and $\theta$; positive velocity components.

## 2. Field equation

Based on the definition of the stream function, $\psi$, velocity components are

$$V_\alpha = \frac{1}{h_\theta} \frac{\partial \psi}{\partial \theta} = \frac{\cosh \alpha + \cos \theta}{R} \frac{\partial \psi}{\partial \theta},\tag{51}$$

$$V_\theta = -\frac{1}{h_\alpha} \frac{\partial \psi}{\partial \alpha} = -\frac{\cosh \alpha + \cos \theta}{R} \frac{\partial \psi}{\partial \alpha}.\tag{52}$$

The field equation for $\psi$ follows from Eq. (4) by substituting the above for $V_\alpha$ and $V_\theta$.

For 2D flows, the vorticity is related to the stream function by:

$$\boldsymbol{\omega} = \hat{\mathbf{e}}_z \nabla^2 \psi \,, \tag{53}$$

where,

$$\nabla^2 \psi = \frac{(\cosh\alpha + \cos\theta)^2}{R^2} \left( \frac{\partial^2 \psi}{\partial \alpha^2} + \frac{\partial^2 \psi}{\partial \theta^2} \right). \tag{54}$$

Hence, the field equation for $\psi$, corresponding to $\nabla^2 \boldsymbol{\omega} = 0$, is

$$\nabla^4 \psi = 0 \,. \tag{55}$$

### 3. *Integration of $\nabla^4 \psi = 0$*

The closed-form solution to $\nabla^4 \psi = 0$ in bipolar coordinates is presented in the following sections.

### (a) *Separation of variables.*

Following the analysis of Jeffery,[22] the biharmonic equation ($\nabla^4 \psi = 0$) may be separated in bipolar coordinates by assuming $\psi$ to be of the form

$$\psi(\alpha, \theta) = (\cosh\alpha + \cos\theta)^{-1} \phi(\alpha, \theta) \,. \tag{56}$$

Substituting Eq. (56) into Eq. (54) and then repeating the $\nabla^2$ operation leads to

$$\frac{\partial^4 \phi}{\partial \alpha^4} + 2 \frac{\partial^4 \phi}{\partial \alpha^2 \partial \theta^2} + \frac{\partial^4 \phi}{\partial \theta^4} - 2 \frac{\partial^2 \phi}{\partial \alpha^2} + 2 \frac{\partial^2 \phi}{\partial \theta^2} + \phi = 0 \,. \tag{57}$$

Assuming the variables are separable, $\phi(\alpha, \theta) = f(\alpha)\, g(\theta)$, Eq. (57) becomes,

$$g^{(4)} + 2(\frac{f^{(2)}}{f} + 1) g^{(2)} + (\frac{f^{(4)}}{f} - 2\frac{f^{(2)}}{f} + 1) g = 0 \,. \tag{58}$$

Since $g$ was assumed to be independent of $\alpha$, the coefficients of $g^{(2)}$ and $g^{(4)}$ must be constants. Setting the coefficient of $g^{(2)}$ equal to $2(1-\tau^2)$,

$$f^{(2)} / f = -\tau^2,\tag{59}$$

where, form of the constant was chosen to obtain eigenfunctions in $\alpha$. The solution of Eq. (59) is

$$f(\alpha) = A(\tau)\sin(\tau\alpha) + B(\tau)\cos(\tau\alpha).\tag{60}$$

Substituting this into the coefficient of $g^{(4)}$ yields

$$\frac{f^{(4)}}{f} - 2\frac{f^{(2)}}{f} + 1 = (\tau^2 + 1)^2.\tag{61}$$

Consequently, the differential equation for $g(\theta)$ is

$$\frac{d^4 g}{d\theta^4} + 2(1-\tau^2)\frac{d^2 g}{d\theta^2} + (\tau^2 + 1)^2 g = 0.\tag{21}$$

Note that this equation in bipolar coordinates is identical to the corresponding equation in toroidal coordinates, Eq. (21). The solution for the stream function is therefore

$$\psi(\alpha,\theta) = (\cosh\alpha + \cos\theta)^{-1}\int_0^\infty \left[ A(\tau)\sin(\tau\alpha) + B(\tau)\cos(\tau\alpha) \right]$$

$$\times \left\{ \sinh(\tau\theta)\left[ C(\tau)\sin\theta + D(\tau)\cos\theta \right] \right.$$

$$\left. + \cosh(\tau\theta)\left[ E(\tau)\sin\theta + F(\tau)\cos\theta \right] \right\} d\tau.\tag{62}$$

*(b) Boundary conditions.*

The forms of the solutions for the 2D and 3D axi-symmetric cases are similar and the boundary conditions are the same. Consequently, the forms of the results for spherical caps [Eqs. (34) - (44)] apply to cylindrical caps as well. Using Eqs. (34) - (44) , the bipolar stream function, Eq. (62), becomes

$$\psi(\alpha, \theta) = (\cosh \alpha + \cos \theta)^{-1} \int_0^\infty K(\theta, \tau) \sin(\tau \alpha) \, d\tau \, . \tag{63}$$

Taking into account that $\sin(\tau \alpha)$ is the eigenfunction of Eq. (63),

$$K(\tau, \theta_c) = \frac{2}{\pi} \int_0^\infty \psi(\alpha, \theta_c)(\cosh \alpha + \cos \theta_c) \sin(\tau \alpha) \, d\alpha \, . \tag{64}$$

Eq. (35) is also valid in bipolar coordinates but the definitions of $V_\alpha$ and $V_\theta$ in terms of the stream function are different. Consequently, the zero shear stress boundary condition yields,

$$\widetilde{K}(\tau, \theta_c) = \frac{2}{\pi} \int_0^\infty \widetilde{\widetilde{\psi}}(\alpha, \theta_c) \sin(\tau \alpha) \, d\alpha \tag{65}$$

in which,

$$\widetilde{\widetilde{\psi}}(\alpha, \theta_c) = -\frac{R}{\cosh \alpha + \cos \theta_c} \frac{\partial}{\partial \alpha} \big[ (\cosh \alpha + \cos \theta_c) V_\theta(\alpha, \theta_c) \big] + \cos \theta_c \, \psi(\alpha, \theta_c) \, . \tag{66}$$

Knowing $K(\tau, \theta_c)$ and $\widetilde{K}(\tau, \theta_c)$, the unknown coefficients, $k_1(\tau)$ and $k_1(\tau)$, are then obtained from Eqs. (39) and (40).

Similar to the spherical cap case, the evaporation rate is related to the variation along the interface of the velocity component normal to the interface, Eq. (45). Using Eqs. (45) and (52), the distribution of the stream function along the surface of the droplet may be written in terms of $V_\theta(\alpha, \theta)\big|_{\theta = \theta_c}$ :

$$\psi(\alpha, \theta_c) = -\int_0^\alpha \frac{R \sinh \alpha'}{\cosh \alpha' + \cos \theta_c} V_\theta(\alpha', \theta_c) \, d\alpha'$$

$$= -\int_0^\alpha \frac{R \sinh \alpha'}{\cosh \alpha' + \cos \theta_c} [J(\alpha') / \rho + V_{\theta,B}(\alpha')] \, d\alpha'. \tag{67}$$

In bipolar coordinates, the components of the velocity follow from Eqs. (51) and (52):

$$R \, V_\alpha(\alpha,\theta) = \sin\theta \, \psi(\alpha,\theta) + \int_0^\infty \frac{\partial K(\theta,\tau)}{\partial \theta} \sin(\tau\alpha) \, d\tau, \tag{68}$$

$$R \, V_\theta(\alpha,\theta) = \sinh\alpha \, \psi(\alpha,\theta) + \int_0^\infty K(\theta,\tau) \cos(\tau\alpha) \, \tau \, d\tau. \tag{69}$$

In cartesian coordinates the horizontal and vertical components of the velocity are:

$$V_x(\alpha,\theta) = \partial\psi / \partial y$$

$$= (\cosh\alpha + \cos\theta)^{-1} \left[ V_\alpha(1 + \cosh\alpha\cos\theta) + V_\theta \sinh\alpha\sin\theta \right], \tag{70}$$

$$V_y(\alpha,\theta) = -\partial\psi / \partial x$$

$$= -(\cosh\alpha + \cos\theta)^{-1} \left[ V_\alpha \sinh\alpha\sin\theta - V_\theta(1 + \cosh\alpha\cos\theta) \right]. \tag{71}$$

## III.  PINNED CONTACT LINE

When the droplet is 'pinned' at the contact line, $R$ is constant and $\theta_c = \theta_c(t)$. This condition will be considered for spherical caps first followed by cylindrical caps. For each of these geometries, two different evaporative fluxes will be considered: uniform and gas-diffusion controlled.

### A.  Spherical cap shape

As reported in our previous paper,[9]

$$\frac{d\theta_c}{dt} = -(2/R)(1+\cos\theta_c)^2 \int_0^\infty \frac{\sinh\alpha}{(\cosh\alpha+\cos\theta_c)^2} \frac{J(\alpha)}{\rho} d\alpha \,,$$ (72)

$$V_\theta(\alpha,\theta_c) = R(\cosh\alpha+\cos\theta_c)^{-1}(d\theta_c/dt) + J(\alpha)/\rho \,,$$ (73)

and

$$\psi(\alpha,\theta_c) = -\frac{d\theta_c}{dt}\frac{R^3}{2}\left[\frac{1}{(1+\cos\theta_c)^2} - \frac{1}{(\cosh\alpha+\cos\theta_c)^2}\right]$$

$$-\int_0^\alpha \frac{R^2\sinh\alpha'}{(\cosh\alpha'+\cos\theta_c)^2} \frac{J(\alpha')}{\rho} d\alpha' \,.$$ (74)

Any physically obtainable distribution of evaporative flux may be considered. Many are possible since different distributions result from different combinations of gas phase velocities and degrees of vacuum into which the evaporation occurs. Two specific cases are considered below.

### 1. Uniform evaporative flux

In this case, the evaporative flux is uniform across the surface of the drop:

$$J(\alpha,\theta_c) = \text{const.} = J_0 \,.$$ (75)

Therefore,

$$\frac{d\theta_c}{dt} = -\frac{2J_0}{R\rho}(1+\cos\theta_c) \,,$$ (76)

$$V_\theta(\alpha,\theta_c) = (J_0/\rho)\left[1-2(1+\cos\theta_c)/(\cosh\alpha+\cos\theta_c)\right],$$ (77)

and

$$\psi(\alpha,\theta_c) = \frac{R^2 J_0}{\rho (\cosh\alpha+\cos\theta_c)}\left(1 - \frac{1+\cos\theta_c}{\cosh\alpha+\cos\theta_c}\right).$$ (78)

Consequently,

$$\widetilde{\psi}(\theta_c, \tau) = (R^2 J_0 / \rho) \left[ -\sqrt{\cosh\alpha + \cos\theta_c} + \frac{\cos\theta_c / 2}{\sqrt{\cosh\alpha + \cos\theta_c}} \right.$$

$$\left. + \frac{4 - (\cos\theta_c - 3)^2 / 4}{(\cosh\alpha + \cos\theta_c)^{3/2}} - \frac{3\sin^2\theta_c(1 + \cos\theta_c)/4}{(\cosh\alpha + \cos\theta_c)^{5/2}} \right]. \quad (79)$$

Using Eqs. (34) and (36) in conjunction with Eq. (A15) of Ref. 9,

$$K(\theta_c, \tau) = -(\sqrt{2}\cosh\pi\tau)^{-1} \left[ 2\tau\cot(\theta_c / 2)\sinh\theta_c\tau + \cosh\theta_c\tau \right] \quad (80)$$

and

$$\widetilde{K}(\theta_c, \tau) = \frac{R^2 J_0}{\rho\sqrt{8}\cosh\pi\tau\sin^2(\theta_c / 2)} \left\{ 2\tau\sinh\theta_c\tau\cot(\theta_c / 2) \right.$$

$$\times \left[ \tau^2(\cos\theta_c - 1) - 2\cos\theta_c - 1 \right]$$

$$\left. + \cosh\theta_c\tau \left[ \tau^2(5\cos\theta_c + 7) - \cos\theta_c + 1 \right] \right\}. \quad (81)$$

## 2. *Diffusive evaporative flux*

A commonly considered flux distribution corresponds to diffusive mass transfer into a stagnant gas (i.e., absent even the gas motion which occurs naturally due to the mass transfer). This model leads to Laplace's equation in toroidal coordinates for the variation of vapor concentration throughout the gas phase. Solving this equation results in the non-physical (singular) evaporative flux at the contact line ($\alpha \to \infty$) for cases of wetting contact angles: $0 \le \theta_c < \pi / 2$.[12] From previous work,[9] it is known that the Gegenbauer transforms exist provided that $\lim_{\alpha\to\infty}[\psi(\alpha, \theta_c)(\cosh\alpha + \cos\theta_c)] = \text{finite}$, and $\lim_{\alpha\to\infty}[\widetilde{\psi}(\alpha, \theta_c) / \sqrt{\sinh\alpha}] = \text{finite}$. These two constraints for obtaining well-behaved liquid-phase transport require the evaporation rate to approach a constant value as $\alpha$ goes to infinity (the contact line). Therefore, the singular behavior of the gas-diffusion solution needs to be remedied to enable uniformly valid, physical solutions to be obtained for the liquid motion.

## B. Cylindrical cap shape

As reported by Petsi and Burganos,[19]

$$\frac{d\theta_c}{dt} = -\frac{\sin^2\theta_c}{R\rho(1-\theta_c\cot\theta_c)}\int_0^\infty \frac{J(\alpha)}{\cosh\alpha+\cos\theta_c}d\alpha\,, \tag{82}$$

$$V_\theta(\alpha,\theta_c) = R(\cosh\alpha+\cos\theta_c)^{-1}(d\theta_c/dt) + J(\alpha)/\rho\,, \tag{83}$$

and

$$\psi(\alpha,\theta_c) = -\frac{d\theta_c}{dt}\frac{R^2}{\sin^2\theta_c}\left\{\frac{\sinh\alpha}{\cosh\alpha+\cos\theta_c} + \cot\theta_c\left[\sin^{-1}\left(\frac{1+\cosh\alpha\,\cos\theta_c}{\cosh\alpha+\cos\theta_c}\right) - \pi/2\right]\right\}$$

$$-\frac{R}{\rho}\int_0^\alpha \frac{J(\alpha')}{\cosh\alpha'+\cos\theta_c}d\alpha'\,. \tag{84}$$

### 1. Uniform evaporative flux

In this case, the evaporative flux is uniform across the surface of the drop:

$$J(\alpha,\theta_c) = \text{const.} = J_0\,. \tag{85}$$

Therefore,

$$\frac{d\theta_c}{dt} = -\frac{J_0}{R\rho}\frac{\theta_c\sin\theta_c}{1-\theta_c\cot\theta_c}\,, \tag{86}$$

$$V_\theta(\alpha,\theta_c) = (J_0/\rho)\left\{1-\theta_c\sin^2\theta_c/\left[(1-\theta_c\cot\theta_c)(\cosh\alpha+\cos\theta_c)\right]\right\}\,, \tag{87}$$

and

$$\psi(\alpha,\theta_c) = \frac{RJ_0}{\rho\,(\sin\theta_c - \theta_c\cos\theta_c)}\left[\frac{\theta_c\sinh\alpha}{\cosh\alpha + \cos\theta_c} + \sin^{-1}\left(\frac{1 + \cosh\alpha\cos\theta_c}{\cosh\alpha + \cos\theta_c}\right) - \pi/2\right].$$

(88)

Consequently,

$$\widetilde{\overline{\psi}}(\theta_c,\tau) = -\frac{RJ_0}{\rho\,(\tan\theta_c - \theta_c)}\left[\frac{\tan\theta_c\sinh\alpha}{\cosh\alpha + \cos\theta_c} + \sin^{-1}\left(\frac{1 + \cosh\alpha\cos\theta_c}{\cosh\alpha + \cos\theta_c}\right) - \pi/2\right]. \quad (89)$$

### 2. Diffusive evaporative flux

Laplace's equation for gas diffusion outside a cylindrical cap drop does not allow an analytical solution when the outer boundary is at infinity since the solution varies logarithmically. Therefore, unlike the 3D axi-symmetric case, a solution does not exist for diffusive evaporative flux in 2D.

## IV. FREELY MOVING CONTACT LINE

When the droplet contact line is free to move (unpinned), the radial distance to the contact line decreases with time $R = R(t)$. One condition previously considered in this case is that the contact angle remains constant during evaporation ($\theta_c = $ constant).[9,19] This will also be assumed here. Since, as discussed above, diffusive evaporative flux does not lead to an acceptable solution, only uniform flux will be considered as the boundary condition in the following sections.

### A. Spherical cap shape

As reported in our previous paper,[9]

$$\frac{dR}{dt} = -\frac{(1+\cos\theta_c)^2}{\sin\theta_c(1+\cos\theta_c/2)} \int_0^\infty \frac{\sinh\alpha}{(\cosh\alpha+\cos\theta_c)^2} \frac{J(\alpha)}{\rho} d\alpha \,, \tag{90}$$

$$V_\theta(\alpha,\theta_c) = \sin\theta_c \cosh\alpha\,(\cosh\alpha+\cos\theta_c)^{-1}(dR/dt) + J(\alpha)/\rho \,, \tag{91}$$

and

$$\psi(\alpha,\theta_c) = -\frac{dR}{dt} R^2 \sin\theta_c \left[\frac{1+\cos\theta_c/2}{(1+\cos\theta_c)^2} - \frac{\cosh\alpha+\cos\theta_c/2}{(\cosh\alpha+\cos\theta_c)^2}\right]$$

$$-\int_0^\alpha \frac{R^2 \sinh\alpha'}{(\cosh\alpha'+\cos\theta_c)^2} \frac{J(\alpha')}{\rho} d\alpha' \,. \tag{92}$$

## 1. *Uniform evaporative flux*

In this case, the evaporative flux is uniform across the surface of the drop:

$$J(\alpha,\theta_c) = \text{const.} = J_0 \,. \tag{93}$$

Therefore,

$$\frac{dR}{dt} = -\frac{J_0}{\rho} \frac{1+\cos\theta_c}{\sin\theta_c(1+\cos\theta_c/2)} \,, \tag{94}$$

$$V_\theta(\alpha,\theta_c) = (J_0/\rho)\left\{1 - \left[(1+\cos\theta_c)\cosh\alpha\right]/\left[(1+\cos\theta_c/2)(\cosh\alpha+\cos\theta_c)\right]\right\} \,, \tag{95}$$

and

$$\psi(\alpha,\theta_c) = -\frac{R^2 J_0 \cos\theta_c}{\rho(\cosh\alpha+\cos\theta_c)(2+\cos\theta_c)}\left(1 - \frac{1+\cos\theta_c}{\cosh\alpha+\cos\theta_c}\right) \,. \tag{96}$$

Consequently,

$$\widetilde{\psi}(\theta_c,\tau) = \frac{R^2 J_0 \cos\theta_c}{\rho(2+\cos\theta_c)}\left[\sqrt{\cosh\alpha+\cos\theta_c} - \frac{\cos\theta_c/2}{\sqrt{\cosh\alpha+\cos\theta_c}}\right.$$

$$\left. -\frac{4-(\cos\theta_c-3)^2/4}{(\cosh\alpha+\cos\theta_c)^{3/2}} + \frac{3\sin^2\theta_c(1+\cos\theta_c)/4}{(\cosh\alpha+\cos\theta_c)^{5/2}}\right] \,. \tag{97}$$

Using Eqs. (80) and (81)

$$K(\theta_c, \tau) = \cos\theta_c \left[\sqrt{2}\,(2+\cos\theta_c)\cosh\pi\tau\right]^{-1}\left[2\tau\cot(\theta_c/2)\sinh\theta_c\tau + \cosh\theta_c\tau\right] \qquad (98)$$

and

$$\widetilde{K}(\theta_c, \tau) = -\frac{R^2 J_0 \cos\theta_c/(2+\cos\theta_c)}{\rho\sqrt{8}\cosh\pi\tau\,\sin^2(\theta_c/2)}$$

$$\times\left\{2\tau\,\sinh\theta_c\tau\,\cot(\theta_c/2)\left[\tau^2(\cos\theta_c-1)-2\cos\theta_c-1\right]\right.$$

$$\left. +\cosh\theta_c\tau\left[\tau^2(5\cos\theta_c+7)-\cos\theta_c+1\right]\right\}. \qquad (99)$$

## B.   Cylindrical cap shape

As reported by Petsi and Burganos,[19]

$$\frac{dR}{dt} = -\frac{\sin^2\theta_c}{\rho(\theta_c-\sin\theta_c\cos\theta_c)}\int_0^\infty\frac{J(\alpha)}{\cosh\alpha+\cos\theta_c}\,d\alpha\,, \qquad (100)$$

$$V_\theta(\alpha, \theta_c) = \sin\theta_c\cosh\alpha\,(\cosh\alpha+\cos\theta_c)^{-1}(dR/dt) + J(\alpha)/\rho\,, \qquad (101)$$

and

$$\psi(\alpha, \theta_c) = \frac{dR}{dt}\frac{R}{\sin^2\theta_c}\left[\frac{\sinh\alpha\sin\theta_c\cos\theta_c}{\cosh\alpha+\cos\theta_c}+\sin^{-1}\left(\frac{1+\cosh\alpha\,\cos\theta_c}{\cosh\alpha+\cos\theta_c}\right)-\pi/2\right]$$

$$-\frac{R}{\rho}\int_0^\alpha\frac{J(\alpha')}{\cosh\alpha'+\cos\theta_c}\,d\alpha'. \qquad (102)$$

### 1.   Uniform evaporative flux

In this case, the evaporative flux is uniform across the surface of the drop:

$$J(\alpha, \theta_c) = \text{const.} = J_0. \qquad (103)$$

Therefore,

$$\frac{dR}{dt} = -\frac{J_0 \, \theta_c \sin \theta_c}{\rho \, (\theta_c - \sin \theta_c \cos \theta_c)} \, , \tag{104}$$

$$V_\theta(\alpha, \theta_c) = (J_0 / \rho) \left\{ 1 - \theta_c \sin^2 \theta_c \cosh \alpha \, / \left[ (\theta_c - \sin \theta_c \cos \theta_c)(\cosh \alpha + \cos \theta_c) \right] \right\}, \tag{105}$$

and

$$\psi(\alpha, \theta_c) = -\frac{J_0 \cos \theta_c}{\rho \, (\theta_c - \sin \theta_c \cos \theta_c)} \cdot$$

$$\times \left[ \frac{\theta_c \sinh \alpha}{\cosh \alpha + \cos \theta_c} + \sin^{-1}\left( \frac{1 + \cosh \alpha \cos \theta_c}{\cosh \alpha + \cos \theta_c} \right) - \pi / 2 \right]. \tag{106}$$

Consequently,

$$\widetilde{\widetilde{\psi}}(\theta_c, \tau) = \frac{R \, J_0 \cos^2 \theta_c}{\rho \, (\theta_c - \sin \theta_c \cos \theta_c)} \left[ \frac{\tan \theta_c \sinh \alpha}{\cosh \alpha + \cos \theta_c} + \sin^{-1}\left( \frac{1 + \cosh \alpha \cos \theta_c}{\cosh \alpha + \cos \theta_c} \right) - \pi / 2 \right].$$

$$\tag{107}$$

## V.  RESULTS AND DISCUSSION

In this section, the velocity distributions for the present viscous flow analysis are compared to the previous distributions obtained for inviscid flow.[9] The cases of pinned and freely moving contact lines are computed for both the uniform evaporative flux and the modified diffusive evaporative flux boundary conditions. Flow patterns are illustrated for $\theta_c < 90^\circ$ (wetting), $\theta_c = 90^\circ$, and $\theta_c > 90^\circ$ (non-wetting).

The results are presented in terms of the dimensionless stream function, $\psi^* = \psi / \psi_0$ and the dimensionless velocity, $V^* = V / V_0$. For axially symmetric flow

$$\psi_0 = R^2 J_0 / \rho \, , \tag{108}$$

and for planar flow,

$$\psi_0 = R^2 J_0 / \rho. \tag{109}$$

For both flows,

$$V_0 = J_0 / \rho. \tag{110}$$

where, $J_0$ is the characteristic evaporative flux. For diffusive evaporation, $J_0$ is defined as

$$J_0 = \rho_g D (Y_s - Y_\infty) / R, \tag{111}$$

where, $\rho_g$ is the density of the vapor-air mixture, $D$ is the binary diffusion coefficient for vapor through air, $Y_s$ is the mass fraction of vapor in the gas phase at the droplet surface (saturation) and $Y_\infty$ is the mass fraction of vapor in the far field.

## A.  Viscous versus inviscid solutions

Figure 3 compares, for the pinned contact, the viscous and inviscid flows within hemispherical, 3(a) - 3(b), and hemicylindrical, 3(c) - (d), drops. For this case ($\theta_c = \pi / 2$) the diffusive flux is uniform for both geometries. It is seen that in both cases streamlines for viscous flow lie farther from the substrate than the corresponding streamlines for inviscid flow. On the other hand, since both viscous and inviscid flows follow the same boundary conditions at the free surface, the flows behave similarly near the droplet surface.

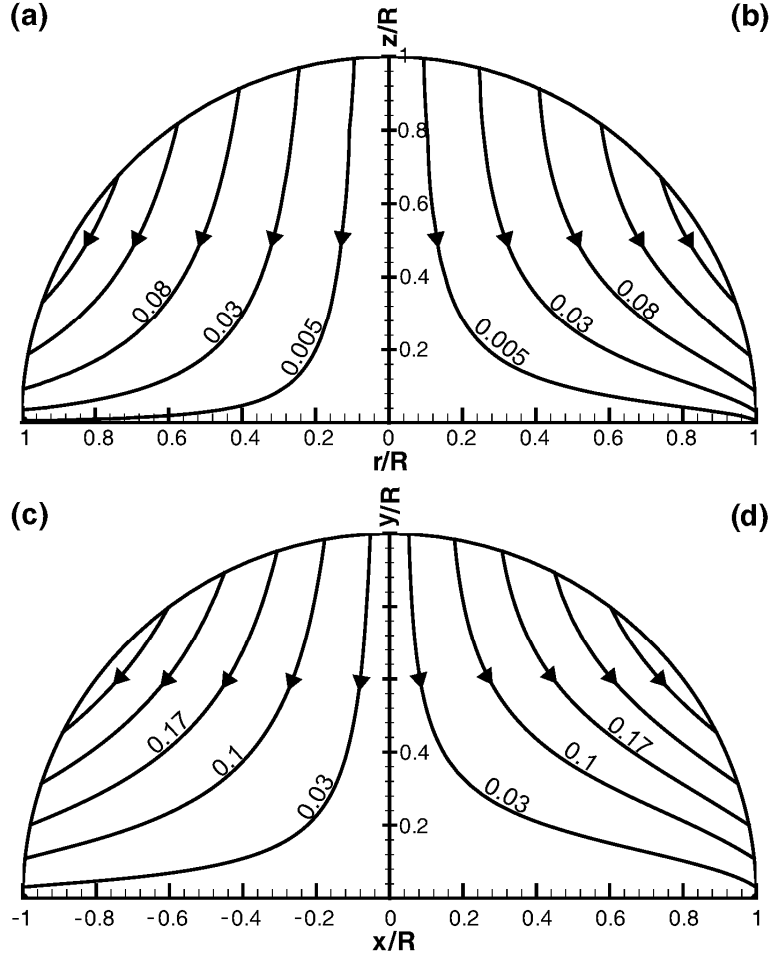

FIG. 3. Contours of nondimensional stream function for pinned contact lines with uniform = diffusive evaporative flux at $\theta_c = 90^\circ$: (a) inviscid, spherical cap, (b) viscous, spherical cap ($\psi_{a,b}^* = 0.005, 0.03, 0.08, 0.15$ and $0.22$), (c) inviscid, cylindrical cap, (d) viscous, cylindrical cap ($\psi_{c,d}^* = 0.03, 0.01, 0.17, 0.24$ and $0.3$).

For the modified diffusive evaporative flux (see Sec. C below), Fig. 4 compares the vertical and radial velocities for inviscid flow to those in viscous flow for pinned contacts with $\theta_c = 40^\circ$. As $r/R$ increases, significant differences are observed in the radial velocities due to the no slip condition at the solid surface. The vertical velocities are qualitatively the same although they differ in magnitude.

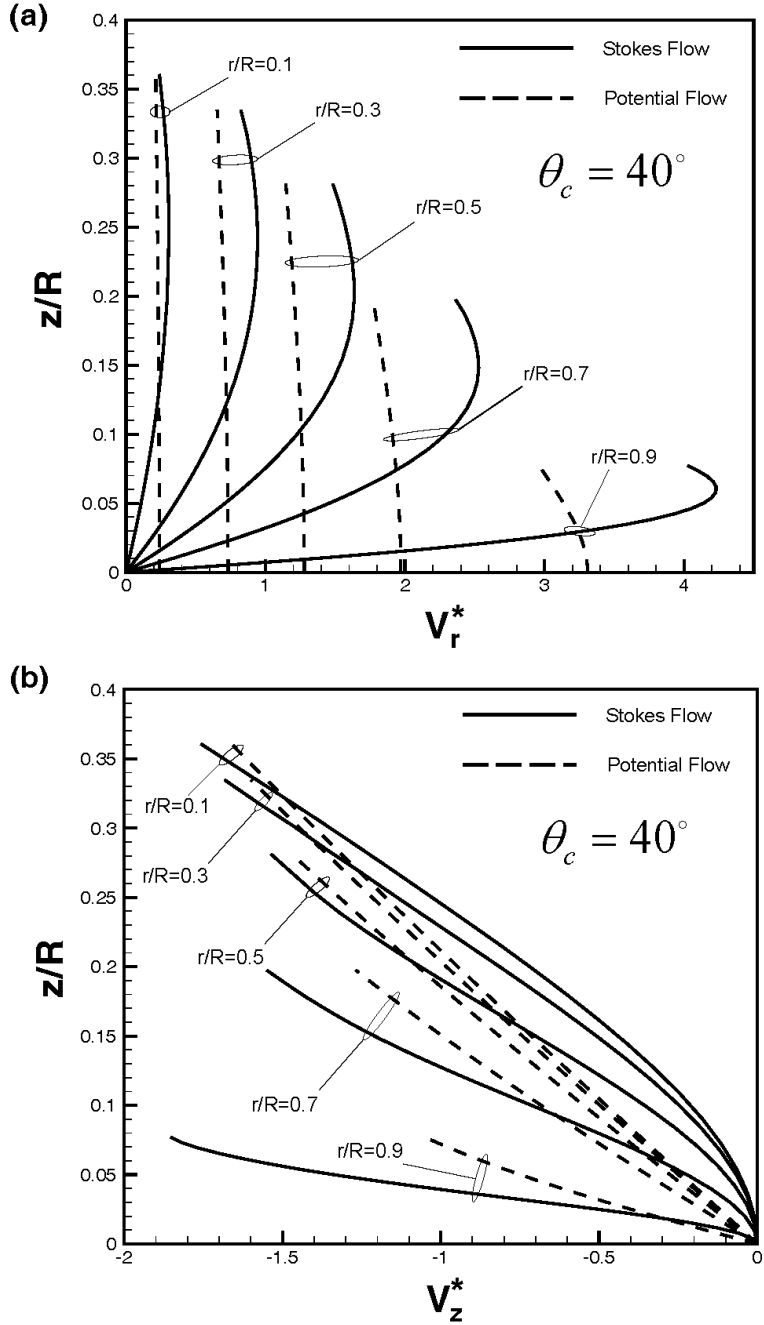

FIG. 4. Nondimensional velocities versus vertical position at different radial positions (r/R= 0.1, 0.3, 0.5, 0.7, 0.9), for a pinned contact line, modified diffusive evaporative flux and $\theta_c = 40^\circ$. The dash lines are inviscid flow;[9] the solid lines are viscous flow. (a) radial velocity, (b) vertical velocity.

In Fig. 4 the contact angle is set to 40° so that the present results could be compared directly with the numerical solution by Hu and Larson.[15] Their solution was only presented graphically. Laying Fig. 4 of this study over Fig. 5 of Ref. 15, showed that the lines were congruent.

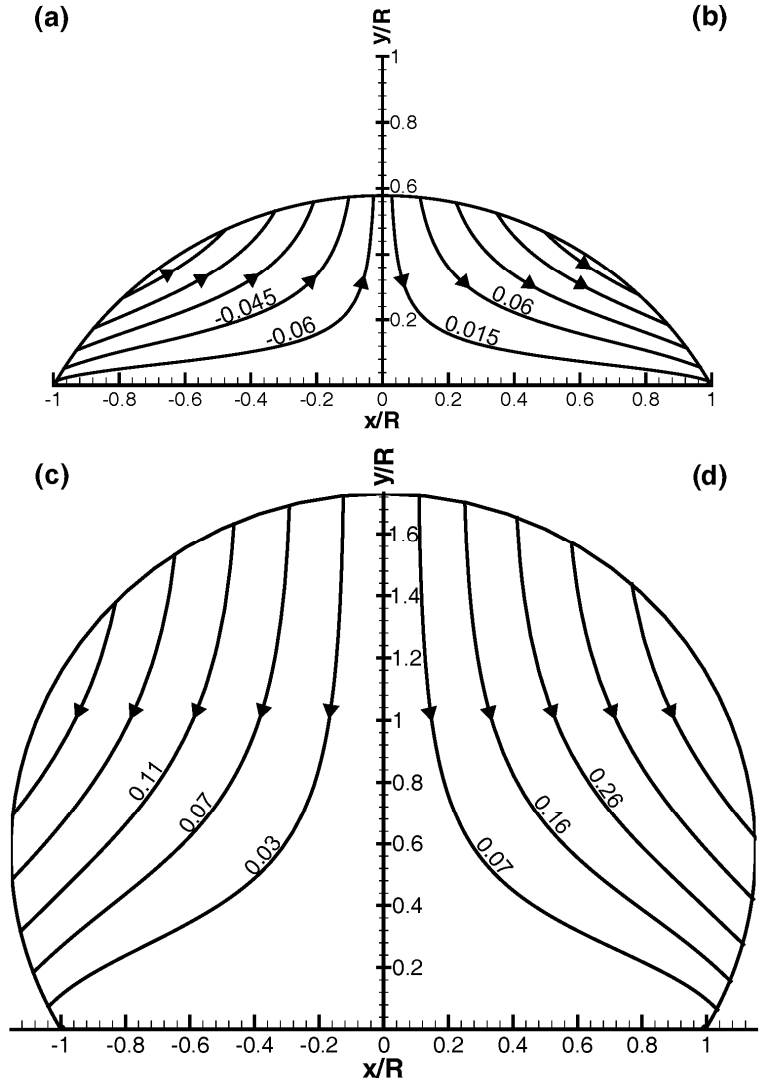

FIG. 5. Contours of nondimensional stream function for viscous flow inside a cylindrical cap exposed to uniform flux (a) $\theta_c = 60^\circ$, freely moving contact line ($\psi^* = -0.06, -0.045, -0.03, -0.015$ and $-0.004$) (b) $\theta_c = 60^\circ$, pinned contact line ($\psi^* = 0.015, 0.06, 0.115, 0.17$ and $0.22$) (c) $\theta_c = 120^\circ$, freely moving contact line ($\psi^* = 0.03, 0.07, 0.11, 0.15$ and $0.185$) (d) $\theta_c = 120^\circ$, pinned contact line ($\psi^* = 0.07, 0.16, 0.26, 0.36$ and $0.46$).

## B. Uniform flux

Since spherical and cylindrical cap drops exhibit the same behavior for various evaporative fluxes, the results for cylindrical cap drops will be used to illustrate the flow structures corresponding to a uniform evaporative flux. The flow patterns computed for different contact angles and contact line conditions are illustrated in Fig. 5. It can be seen that when the contact line is pinned, the flow is from the center of the drop to its edge. On the other hand, when the contact line is free to move, distinctively different flow patterns are observed for wetting ($\theta_c < \pi/2$) and non-wetting ($\theta_c > \pi/2$) conditions – see Figs. 5(a) and 5(c). (This is consistent with the behavior in inviscid flow.[9, 18])

## C.   Modified diffusive evaporative flux

It is shown in Ref. 9 that: (i) for $\theta_c > 90^o$, the evaporative flux decays to zero as $(r/R) \to 1$, (ii) for $\theta_c = 90^o$, the evaporation rate is uniform over the free surface, and (iii) for $\theta_c < 90^o$, the evaporation rate diverges at the contact line. In order to prevent the non-physical divergence when $\theta_c < 90^o$, Fischer[14] assumed here that the evaporative flux decays exponentially near the contact line ($\alpha > 7$). We have used that assumption in the following calculations when $\theta_c < 90^o$ and refer to the boundary condition for these cases as *modified diffusive evaporative flux.*

Figure 6 presents the four cases considered: $\theta_c = 60^o, 120^o$ for both pinned and freely moving contact lines. When the contact line is pinned, the flow is directed from the center of the drop to its edges ('coffee-ring' phenomenon). The flow behavior remains the same even for contact angles greater than ninety degrees where the evaporative flux distribution is quite different. On the other hand, when the contact line is freely moving the flow pattern is more intricate. In a given drop, fluid flows both toward and away from the edge, making it unlikely that a colloidal drop would deposit a 'coffee ring' pattern of particles.

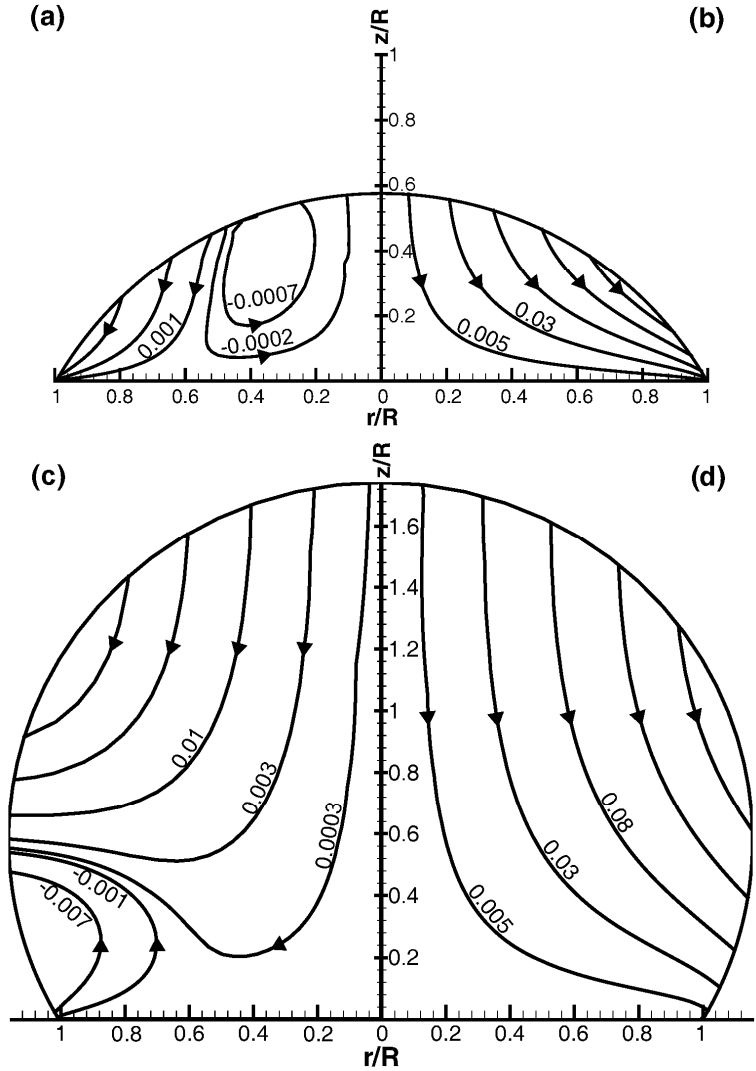

FIG. 6. Contours of nondimensional stream function for viscous flow inside a spherical cap exposed to *modified diffusive evaporative flux* (a) $\theta_c = 60^o$, freely moving contact line ($\psi^* = $ -0.0007, -0.0002, 0.001, 0.006 and 0.02) (b) $\theta_c = 60^o$, pinned contact line ($\psi^* = $ 0.005, 0.03, 0.08, 0.15 and 0.22) (c) $\theta_c = 120^o$, freely moving contact line ($\psi^* = $ -0.007, -0.001, 0.0003, 0.003 , 0.01, 0.02 and 0.03) (d) $\theta_c = 120^o$, pinned contact line ($\psi^* = $ 0.005, 0.03, 0.08, 0.15 and 0.22).

# ACKNOWLEDGMENT

The importance of this problem was brought to our attention by Prof. R.C. Wetherhold.